\documentclass[sigconf]{acmart}
\acmConference[ICSE 2024]{46th International Conference on Software Engineering}{April 2024}{Lisbon, Portugal}

\usepackage{listings}
\usepackage{xcolor}
\usepackage{listings-rust}
\usepackage{todonotes}
\usepackage{times}

\usepackage{soul}
\usepackage{url}
\usepackage{ulem}
\usepackage{latexsym}
\usepackage{mathrsfs}
\usepackage{alltt}
\usepackage{todonotes}
\usepackage{pifont}
\usepackage{graphicx}
\usepackage{color,xcolor}
\usepackage{tcolorbox}
\usepackage{braket}
\usepackage{bbm}
\usepackage{hyperref}
\usepackage{subfig}
\usepackage{balance}
\usepackage{algorithm}
\usepackage{algpseudocode}
\usepackage{balance}
\usepackage{enumerate}
\usepackage[shortlabels]{enumitem}

\lstdefinestyle{icse-style}{
  backgroundcolor=\color{white},     % Background color
  basicstyle=\small\ttfamily,        % Basic font style
  breakatwhitespace=false,           % Break at white space only
  breaklines=true,                   % Automatically break lines
  captionpos=b,                      % Caption position: bottom
  commentstyle=\color{gray},         % Comment style
  frame=single,                      % Frame around code
  keywordstyle=\color{blue},         % Keyword style
  language=Java,                     % Language
  numbers=left,                      % Line numbers
  numbersep=5pt,                     % Distance of line numbers from code
  numberstyle=\tiny\color{gray},     % Line number style
  showspaces=false,                  % Don't show spaces
  showstringspaces=false,            % Don't show spaces in strings
  showtabs=false,                    % Don't show tabs
  stringstyle=\color{orange},        % String style
  tabsize=2                          % Tab size
}

\def\cratesioProjects{94715}
\def\cratesioNumWarnings{8,095,259}
\def\cratesioLOC{374,424,742 LOC}

\def\cratesioDensityKLOC{21}

\def\rustlangProjects{70}

\AtBeginDocument{%
  \providecommand\BibTeX{{%
    \normalfont B\kern-0.5em{\scshape i\kern-0.25em b}\kern-0.8em\TeX}}}

\setcopyright{acmcopyright}
\copyrightyear{2018}
\acmYear{2018}
\acmDOI{XXXXXXX.XXXXXXX}

\acmConference[Conference acronym 'XX]{Make sure to enter the correct
  conference title from your rights confirmation emai}{June 03--05,
  2018}{Woodstock, NY}
\acmPrice{15.00}
\acmISBN{978-1-4503-XXXX-X/18/06}

\begin{document}

\title{Unleashing the Power of Clippy in Real-World Rust Projects}
% \title{Fixing High-Frequency Lints in Rust Projects}


\author{Chunmiao Li}
\affiliation{%
  \institution{Huawei Beijing Research Center}
  \country{China}}
\email{chunmiaoli1993@gmail.com}

\author{Yijun Yu}
\affiliation{%
	\institution{The Open University}
	\country{UK}}
\email{y.yu@open.ac.uk}

\author{Haitao Wu}
\affiliation{%
	\institution{Huawei Waterloo Research Center}
	\country{Canada}}
\email{haitao.wu@huawei.com}

\author{Luca Carlig}
\affiliation{%
	\institution{Huawei Ireland Research Center}
	\country{Ireland}}
\email{luca.carlig@huawei.com}

\author{Shijie Nie}
\affiliation{%
	\institution{Fujitsu R\&D Center}
	\country{China}}
\email{nieshijie@fujitsu.com}

\author{Lingxiao Jiang}
\affiliation{%
	\institution{Singapore Management University}
	\country{Singapore}}
\email{lxjiang@smu.edu.sg}

\renewcommand{\shortauthors}{Trovato and Tobin, et al.}

\begin{abstract}
Clippy lints are considered as essential tools for Rust developers, as they can be configured as gate-keeping rules for a Rust project during continuous integration. Despite their availability, little was known about practical application and cost-effectiveness of the lints in reducing
code quality issues. In this study, we embark on a comprehensive analysis to unveil the true impact of Clippy lints in the Rust development landscape. 
%
% Clippy lints have been available for Rust developers to configure their projects as gate-keeping rules for continuous integration. However, it was not known whether they are applied, and if so, which ones are cost-effective in reducing manual effort to fix the warnings. 
%
The study is structured around three interrelated components, each contributing to the overall effectiveness of Clippy. 
Firstly, we conduct a comprehensive analysis of Clippy lints in all idiomatic crates-io Rust projects with an average warning density of 21/KLOC.
The analysis identifies the most cost-effective lint fixes, offering valuable opportunities for optimizing code quality. 
Secondly, we actively engage Rust developers through a user survey to garner invaluable feedback on their experiences with Clippy. User insights shed light on two crucial concerns: the prevalence of false positives in warnings and the need for auto-fix support for most warnings.
% Understanding these challenges provides invaluable guidance for refining Clippy's performance and usability.
Thirdly, building upon these findings, we engineer three innovative automated refactoring techniques to effectively fix the four most frequent Clippy lints. As a result, the warning density in Rosetta benchmarks has significantly decreased from 195/KLOC to an impressive 18/KLOC, already lower than the average density of the crates-io Rust projects.
These
%outstanding progress
results demonstrate tangible benefit and impact of our efforts in enhancing the overall code quality and maintainability for Rust developers.

% In this paper, we first analyse the landscape of Clippy lints in all idiomatic crates-io Rust projects with a warning density of 21/KLOC and identify which lint fixes would be the most cost-effective; then 
% we conduct a user study to learn the necessary steps to adopt Clippy as gate-keeping rules; and 

% finally, we engineer three automated refactoring techniques to fix the four most frequent lints, bringing warning density of Rosetta benchmarks from 195/KLOC down to 18/KLOC. This density is already lower than the average of the crates-io Rust projects. 
\end{abstract}
%% The author(s) should pick words that accurately describe


\maketitle

\section{Introduction}
% This paragraph can be removed to save space
%Voted as the most favored programming language according to the Stack Overflow surveys since 2015~\footnote{https://insights.stackoverflow.com/survey/2021\#overview}, Rust has garnered significant attention from both academia and industry. Researchers are particularly interested in Rust's safety claims, and have applied rigorous proofs to validate them~\cite{jung2017rustbelt,matsushita2021rusthorn,astrauskas2022prusti}. Having been adopted for developing safety-critical artifacts, Rust projects often see impressive gains in efficiency~\cite{levy2017multiprogramming, parity, redox, servo}.

%Whether a novice programmer or an experienced developer but new to Rust, 
The error messages generated by the Rust compiler (\textsf{rustc}) are useful for developers to identify and diagnose suspicious code segments \cite{googleInsight,abtahi2020learning}. 
Complementing the compiler, linters can also play an important role in promoting the adherence to certain coding style conventions and best practices. Prominent linters utilized in the Rust ecosystem include Clippy~\cite{clippy}, Rustfmt~\cite{rustfmt}, and Cargo-Geiger~\cite{cargo-geiger}. Among them, the Rust community particularly emphasizes on the importance of heeding the warnings provided by Clippy to mitigate common errors and promote the adoption of idiomatic conventions. Clippy provides a set of more than 600 lints in addition to the built-in \textsf{rustc} lints. These lints are divided into nine distinct categories that address correctness and style aspects. Each category is assigned a default lint level, namely Allow, Warn, or Deny, indicating the severity with which the lints are reported. 

Nevertheless, there is a conspicuous \textbf{absence} of research examining the impact of Clippy in real-world Rust projects, despite similar studies being conducted for other programming languages like Python~\cite{oliveira2022lint, simmons2020large, bafatakis2019python} and JavaScript~\cite{kavaler2019tool}. Consequently, the distribution of warnings and the prevalence of specific warnings in idiomatic projects remain unclear. It is important to emphasize that the adoption of varying conventions by different projects may lead to the selection of distinct subsets of available lint rules for gate-keeping purposes. Therefore, it is imperative to undertake such studies to provide valuable guidance and insights for developers, ultimately improving code quality and best practices within the Rust ecosystem.

Moreover, investigating developers' perceptions and responses to Clippy's warnings would be invaluable in comprehending the practical application of Clippy and promoting its widespread adoption. Similar studies~\cite{ayewah2008report,johnson2013don,layman2007toward,tomasdottir2018adoption,tomasdottir2017and} conducted for other linters have provided helpful insights, but 
\textbf{none} have specifically addressed how developers evaluate the effectiveness of Clippy's warnings in their projects and how they treat these warnings. Additionally, the types of Clippy warnings that developers typically configure or prioritize in their projects remain uncertain. Understanding these preferences and configurations is crucial in optimizing Clippy's impact and catering to developers' specific needs and coding conventions. 

% Furthermore, the reported lints often exhibit significant overlap with errors or warnings commonly encountered in other programming languages, such as C. 

% incorporating all lints universally without verification poses overheads on and uncertainty 

%Nonetheless, 

% Considering that each programming language possesses its distinct characteristics, tools, and coding style culture, it holds value to undertake a comprehensive evaluation of linting practices within the context of Rust. Surprisingly, a dearth of existing research investigating linting in the Rust is evident.

Lastly, exploring lint violations and their resolutions is not an unfamiliar research direction~\cite{johnson2013don,tomasdottir2017and,habchi2018adopting,rafnsson2020fixing,tomasdottir2018adoption,christakis2016developers}, it is interesting to study the measures to fix Clippy warnings. While Clippy is more capable of identifying and raising alerts for security threats than \textsf{rustc}, its reports may not be accurate or complete.
For example, the program depicted in Lines 1-5 of Figure~\ref{fig:example} reads the contents of a file named \textsf{fpath} and prints them. Although the program can be successfully compiled and executed by \textsf{rustc} using \texttt{cargo run}, enabling the clippy option \texttt{-W clippy:: unwrap\_used} would raise a warning in Lines 9-12 regarding the usage of \textsf{unwrap()}\footnote{We observed that this lint is predominantly configured in idiomatic projects. We will present the detailed results in subsequent sections.}. Clippy suggests replacing {\sf unwrap()} with {\sf expect()}; however, this solution fails to address the underlying issue. As evidenced in Lines 16-19, even though the \textsf{unwrap\_used} warning disappears after the changes, another warning \textsf{expect\_used} complains that if the function \texttt{read\_to\_string()} on Line 3 fails to return a \textsf{String}, the program would prematurely halt due to a panic. To resolve both potential warnings, one has to employ the \textsf{if let} feature of Rust's control flow syntax in Lines 24-27. By incorporating this modification, the program can exit gracefully even in the event of a failed file read operation.
\begin{figure}[h]
\begin{lstlisting}[style=icse-style,escapechar=\%]
use std::fs::read_to_string;
fn main() {
  let contents = read_to_string("fpath").unwrap();
  println!("{contents}");
}

// diagnostics by 
//  `cargo clippy -- -W clippy::unwrap_used`
warning: used `unwrap()` on a `Result` value
 --> src/main.rs:3:20
3 let contents = read_to_string("fpath").unwrap();
  = help: ... consider using `expect()` ...
  
// suggested fix and further diagnostics by
//  `cargo clippy -- -W clippy::expect-used`:
warning: used `expect()` on a `Result` value
 --> src/main.rs:3:20
3 let contents = read_to_string("fpath").expect("not found");
  = help: if this value is an `Err`, it will panic
  
// manual fix and no warnings reported by 
//  `cargo clippy -- -W clippy::expect-used 
//         -W clippy::unwrap-used`
fn main() {
  if let Ok(contents) = read_to_string("fpath")
    { println!("{contents}");}
}
\end{lstlisting}
\caption{An example of `clippy` warning and a manual fix\label{fig:example}}
\end{figure}
% start: 7/21 ipad notes
% Why unwrap_used? 我的解释是可以的 因为后面统计用的比较多
% The official document for the \textsf{unwrap}~\footnote{https\://rust-lang.github.io/rust-clippy/master/index.html\#/unwrap\_used} lint 
% illustrates that explains that although using \textsf{unwrap} can be a convenient choice for quick-and-dirty code, it is advisable to handle the None or Err case more explicitly.
In addition to detecting such warnings, for those lints that are tagged {\sf MachineApplicable}, Clippy could provide an auto-fix option by entering the command {\tt cargo clippy --fix}.
%However, the other lints cannot be found\todo{unclear what this sentence means} automatically. 
However, many other lints cannot be fixed automatically.

% \todo{"these problems" may be too vague; may explicitly write out our high-level research questions for the study before this paragraph}, 
To address these unresolved problems, we first conducted a landscape study to analyze the distribution of Clippy warnings across all projects hosted at crates.io, along with a longitudinal survey examining the evolution of warnings in four representative official Rust projects. Our observations revealed a highly skewed distribution of warnings, with over 50\% of warnings attributed to the top-1 lints, and over 80\% of warnings identified by the top-5 lints. This raised uncertainty regarding the automatic resolution of these high-frequency warnings. Furthermore, we observed that certain projects clearly adopt gate-keeping practices, incorporating Clippy as part of their CI/CD process to ensure code with minimal warnings before committing, while others do not follow such practices. Consequently, the rationale behind adopting Clippy as a gate-keeping step, particularly concerning its auto-fix capabilities, remained ambiguous.
 
% Additionally, some projects clearly adopts gate-keeping practices, i.e., using Clippy as part of the CI/CD process to ensure code with few warnings can be committed, while others do not; hence it was still unclear whether the decision to adopt Clippy as a gate-keeping step is due to its capability of auto-fixes.
% \item the Clippy auto-fixes are often counter-productive, even when they can be applied developers often need to undo the changes; hence it was unclear whether developers would like to disable certain Clippy lints due to such limitations.
% \end{itemize}

Following on these observations, we performed a user survey on the usage of Clippy in real-world Rust projects. The responses revealed that Clippy is  indeed regarded as a valuable linter. However, Clippy is frequently not selected as the gate-keeper in continuous integration due to the high false positives and limited automated fixes for most warnings.

Finally, to foster the auto-fix capabilities for Clippy warnings, we proposed three promising strategies
% address the limitations of Clippy's auto-fix functionality 
through the combination of rule transformation, direct integration of fix operations into Clippy, and the development of specialized shell script. They offer effective means for developers to remove warnings and enhance the overall quality and reliability of Rust projects. It is worth noting that, to the best of our knowledge, we are the \textbf{first} to tackle this specific issue. As a result, the warning density in Rosetta benchmarks has significantly decreased from 195/KLOC to an impressive 18/KLOC, already lower than the average density (21/KLOC) of the crates-io Rust projects.

% We evaluated the feasibility of automating the fixes for high-frequency lints. 

% Informed by the high-frequency lints of Rosetta Code benchmarks, which is largely consistent with those found in Crates-IO, 
% \begin{itemize}[leftmargin=1em]
% \item 
% \item 
% evaluated the Rust projects after remove the high-frequency warnings thanks to the auto-fix measures in terms of warning density in the benchmarks in our datasets. 
% \end{itemize}

From these studies, we reveal new implications to three beneficiaries. Firstly, we suggest Rust developers consider incorporating Clippy into their projects for its adherence to optimal practices and style guides. Secondly, it is an opportunity for researchers to explore novel techniques to assist users in effectively resolving warn-
ings, with particular attention to addressing the most frequently occurring warnings as we identified. Thirdly, creators of Clippy lints might consider reduce false positives and provide organized, illustrative examples to boost user satisfaction and usability.

% we observed that Clippy lints are not widely adopted as gate-keeping for average Rust projects because their are false positives and the existing auto fixes of the lints rarely reduce the warning density. Using a benchmark, we also demonstrated that auto-fix the high-frequency lints is not only possible, but also effective when applied in reducing the warning density. 

Our contributions are as follows:
\begin{itemize}
    \item We conducted the {\tt first} study of Clippy's impact to real-world Rust projects, unleashing the distribution of warnings and revealing user feedback to Clippy.
    \item We proposed three effective solutions to fix Clippy warnings that cannot be automatically removed using its auto-fix functionality.
    \item We collected cost-effective lints and users expectations to Clippy, aiming to foster its widespread adoption to enhance the safety and idiomaticity of Rust programs.
\end{itemize}

The subsequent sections of this paper are structured as follows: Section 2 presents background of linters and code transformation; Section 3 reports our landscape study of Rust projects hosted on the Crates-IO and longitudinal findings on four official typical projects; Section 4 outlines a user survey to gain their feedback concerning Clippy; Section 5 demonstrates the effectiveness of proposed three approaches to reduce the high-frequency Clippy warnings; Section 6 discusses potential threats to the validity of our work. Section 7 provides related work and Section 8 concludes the paper.

\section{Background}

%\subsection{Diagnostics in Software Engineering}
% This section provides an overview of error diagnostics in software engineering, including how compilers expose warnings as a guideline for developers to fix errors and program repair techniques to automatically fix errors.

\subsection{Linters}
In order to encourage developers to produce more idiomatic programs and enhance overall code quality, linters employ diagnostic warnings, to highlight potential code smells or linting rule violations. These warnings typically indicate the location of suspicious code snippets and the associated types of potential issues. Different programming languages are accompanied by various linters designed to identify specific problems. For instance, Bandit~\cite{bandit} is utilized to detect common security issues in Python code, while pep8 checks Python code against established style conventions. Similarly, for JavaScript, popularly used linters include ESLint~\cite{eslint}, Jshint~\cite{jshint}, and Jslint~\cite{jslint}. Within the Rust ecosystem, several prominent linters are employed, such as Clippy~\cite{clippy}, Rustfmt~\cite{rustfmt}, and Cargo-Geiger~\cite{cargo-geiger}. Among these, the Rust community predominantly advocates the consideration of warnings emitted by Clippy~\cite{clippy} to steer clear of common errors and adhere to idiomatic conventions. Notably, Clippy encompasses a set of lints that are not available in the default Rust compiler lint list, thus providing additional opportunities for code analysis and improvement.

% Meanwhile, compilers may also generate extra repair suggestions to developers or offer several options to automatically refactor the warning away from the code.
% Industrial experience reports have confirmed the value of fixing these warnings on code quality and security assurance: for example, Microsoft discovered potential threats in code review process from compiler warnings~\cite{howard2006process}. 

% Currently\footnote{2023/8/1}, more than 650 Clippy lints are classified into nine categories (like  Clippy::style and Clippy::correctness) with different levels (i.e., warn/deny/allow). For example, the {\tt unwrap\_used}~\cite{unwrapUsed} lint rule is allowed by default, which checks for any {\tt .unwrap()} or {\tt .unwrap\_err()} calls in Rust programs. 

% In addition to showcase the unwrap used examples by replacing it with {\tt expect("more helpful message")} or by a question mark macro {\tt `?'}. 
%In addition to Clippy, other linters are also proposed~\cite{}

% \begin{table}[h]
% \begin{tabular}{|l|l|l|}
% \hline
%              & Functionality & GitHub Stars \\ \hline
% Clippy       &               &              \\ \hline
% rustfmt      &               &              \\ \hline
% cargo-geiger &               &              \\ \hline
% MIRAI        &               &              \\ \hline
% dylint       &               &              \\ \hline
% \end{tabular}
% \end{table}

\subsection{Code Transformations}
% A program transformation takes a computer program as input and generates
% another program as output. In a narrow perspective, source-to-source code
% transformation is a kind of program transformations, which converts one kind of
% high-level language (i.e. source code) to another kind of high level language.
% Txl is a programming language named designed to support such code
% transformations, which is chosen in this paper for the following reasons~\cite{Cordy06ACM-STAGTXL}: 

Source-to-source code transformation converts high-level source code into another high-level language. To support such transformations, we choose Txl, a programming language designed with specific advantages. Txl's fidelity ensures deterministic and consistent processing by tokenizing and parsing input code into a parse tree based on a grammar definition. Additionally, Txl's user-oriented grammar enables efficient transformations at the token-tree level, maintaining speed in the process. Moreover, Txl's scalability allows for convenient updates to the grammar, making it well-suited for custom scenarios and rapid prototyping, particularly in languages like Rust~\cite{Cordy06ACM-STAGTXL, Thurston06SCAM-TXL}.

% \begin{itemize}[leftmargin=1em]

% \item Fidelity. Txl tokenises input (original code) and parses it into a tree
% according to the grammar definition specified in a notation similar to
% extended Backus-Naur Form (eBNF). All following operations are based on the
% parse tree, therefore, the grammar definition guarantees the whole procedure
% deterministic and consistent.

% \item Speed. Using a more ``natural" user-oriented grammar to describe input
% structures rather than compiler-style ``implementation" grammars, Txl generates
% transformations that work on the token-tree level, maintaining transformation
% efficiency.

% \item Scalability. For customised scenarios or rapid prototyping programming
% language such as Rust, adjustable grammar definition and rule-based
% operations of Txl make updates of the grammar incremental and convenient~\cite{Thurston06SCAM-TXL}. 

% \end{itemize}

\makeatletter
\@addtoreset{research}{section}
\makeatother

\section{Detecting Diagnostic Warnings from Rust Landscape}
% ii) Rigor: The soundness, clarity and depth of a technical or theoretical contribution, and the level of thoroughness and completeness of an evaluation.
% iv) Verifiability and Transparency: The extent to which the paper includes sufficient information to understand how an innovation works; to understand how data was obtained, analyzed, and interpreted; and how the paper supports independent verification or replication of the paper’s claimed contributions. Any artifacts attached to or linked from the paper may be checked by one reviewer.

% We want to answer the following questions:

% Examine the utilization of Clippy within the top 20 GitHub projects, ranked by the highest number of stars as of the present moment:

% Conduct a similar analysis on other Rust linters, as mentioned in question 1, but with the option to selectively include specific entries:

% Clippy warning types, code snippet in idiomatic rust projects

% \subsection{}

% We want to extract the practical modification cases that developers manually fixed Clippy warnings, whereas there are some challenges exist. Firstly, after surveyed the commits records on GitHub, we found that some warnings are turned on as \textbf{allow} instead being fixed actually, so that the cases that 

% Which types of warnings and errors reported by Rust's linter are relatively unique and prevalent compared to other languages such as C/C++?

% \todo{NJU students}

% \subsection{Detecting Diagnostic Warnings from Rust Landscape}

{\bf RQ1.} {\it How many warnings exist in the Rust development landscape?  How are they distributed over warning types? Are they fixed by developers?}

Although the use of Rust has potentially infinite projects, the
Crates-IO\cite{cratesIO} offers a holistic hosting service to most
of widely used and downloadable Rust projects, which is well over \cratesioProjects{}. 
Meanwhile, the Rust official organisation hosts on the GitHub rust-lang%\footnote{https://github.com/rust-lang} 
over 78 Rust projects, marked by the major programming language of artefacts. 
%Therefore, we are interested to know how dense are the warnings to compare community projects with official projects. Furthermore, we have developed a few Rust projects for research, which may also offers a comparison with the landscape. 

% 上次reviewer没有complain的部分 
% 用chatgpt润色一下这一节
\subsection{Our Approach}
We developed a tool to detect warning messages of types. For each warning message we found its location between the starting and the ending offsets in the source file. There are also hints by the lint rules on how to fix the detected warning.
%These are exemplified by the two warnings in Lines 9-12 and Lines 16-19 in Figure~\ref{fig:example}.

Using this lint tool, we can check a landscape of Rust projects which includes
\cratesioProjects{} Rust projects from Crates-IO, the default site hosting
open-source Rust libraries. These also include some of the artefacts built from the \rustlangProjects{} Rust projects officially hosted on GitHub %~\footnote{Time:} 
that defines the language and fundamental tool chains. From this survey, we can obtain the total number of warnings, and the measure
of KLOC for a Rust project, from which we can derive a density metric of
warnings as the number of warnings per KLOC. 
% can be seen as candidate projects where Clippy has been used as a gate keeper for their maintainers. 

\subsection{Evaluation Results}
Since Clippy is an open-source project, there are already over 600 lint rules to check. Of course, not all of them are relevant to development and not all of them are turned on as the default choice of project teams. In fact, every project can configure a subset of these rules as Rust compiler switches. For an unbiased data extraction, therefore, we shall use individual Rust project's own configuration to extract the warnings the team cares about. 

\subsubsection{Landscape Survey}

\paragraph{Crates-IO} To study the current landscape of warnings, it is
impractical to look into the past history due the large number of revisions per
project. However, it is sufficient to do the survey on the latest version of
the projects. Following the semantic versioning (semver) convention of Rust
community, a version x.y.z of a crate library is considered more recent by
lexical ordering of x, y and z. In other words, x.y.z is later than $x-1$.*.*,
or x.$y-1$.*, or x.y.$z-1$. 

As of the experiments, there are \cratesioProjects{} projects on Crates-IO,
which has \cratesioLOC{} for their latest release, resulting in 484 types of
clippy warnings used in various configurations.

Altogether we found \cratesioNumWarnings{} instances of 403 types of warnings
actually are still with these latest releases. The distribution of these
warnings is long-tailed: 
over 113 types have fewer than 10 warnings, while the top of them have got much higher numbers. 

Figure~\ref{fig:cratesio-warnings} shows the top 20 types of warnings. (The detailed data is in the supplementary.)
The top one gets 4,887,114, which is already over 50\% of all warnings. On the
other hand, not all of these warnings can be fixed automatically and not all
Rust programmers would agree on each other's choice. 
The largest category {\tt default\_numeric\_fallback}, e.g., requires to
rewrite constant numbers e.g., {\tt 0} as {\tt 0\_i32} with an explicit data
type {\tt i32}. It is not always necessary because the Rust compiler could
infer such types most of the time. Nonetheless we have to include it in our
statistics as long as the project teams decide to do so. 
% comment out for the sake of automated code transformations
% A fix to such a warning does not require sophisticated code transformations.
\begin{figure}[th]
\includegraphics[width=\columnwidth]{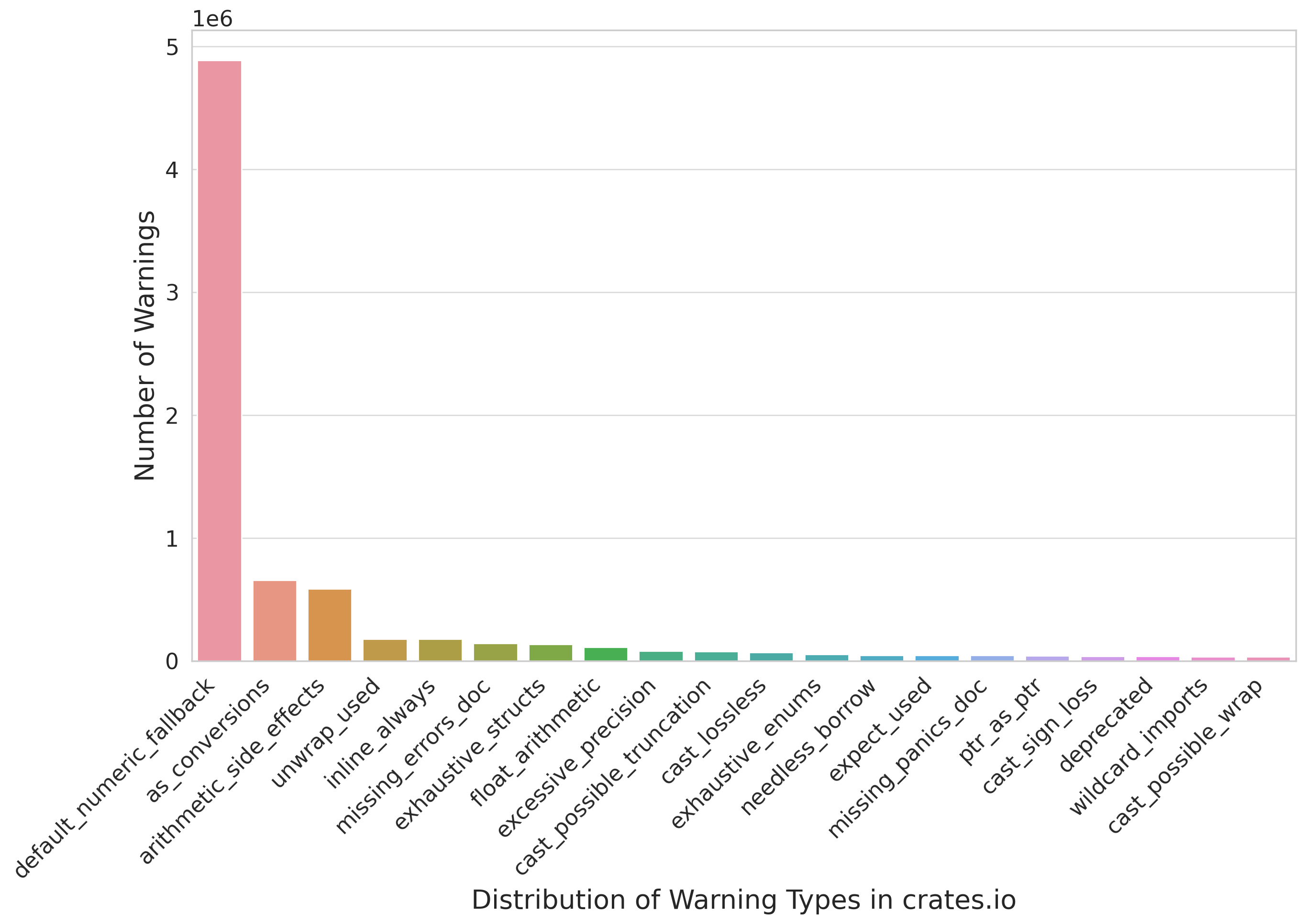} 
\caption{Top 20 types of warnings in Rust projects (crates.io)
	%{\tt unwrap\_used} ranks No. 4 
	\label{fig:cratesio-warnings}}
\end{figure}

On average, \cratesioDensityKLOC{} per KLOC is the density of warnings for the
Crates-IO projects.  %Amongst them, the {\tt unwrap\_used} rule ranks no. 4 high frequency in the occurrences. 
The distribution of warnings is highly skewed, with over 50\% of warnings found by the top-1 lints, and over 80\% of warnings found by the top-5 lints; hence it was still unclear whether these high-frequency warnings can be addressed automatically. 

\subsubsection{Longitudinal Survey}

To understand how the quality of projects evolve along the history of its
repository, it is useful to count the warnings on every revisions. The
changes of warning counts could indicate whether maintenance effort has
been spent on reducing the diagnostic warnings, and to understand whether such
control is as tight as gate-keeping. However, studying the evolution of warnings in all projects on crates.io is time-consuming. Therefore, we focused our study on the evolution of warnings in four typical Rust projects from the official rust-lang organization: RustFix~\cite{rustfix}, Log~\cite{log}, Git2-Rs~\cite{git2-rs}, and CC-Rs~\cite{cc-rs}. RustFix is an official Rust library project designed to fix errors. Log is another official Rust library project used for managing logs. Git2-Rs, on the other hand, is a Rust library project that interfaces with the popular libgit2 library. CC-Rs is a project responsible for compiling C/C++ code into shared libraries for interaction with Rust.

\begin{figure*}[ht]
 
\begin{tabular}{cc}
 \includegraphics[width=0.45\textwidth]{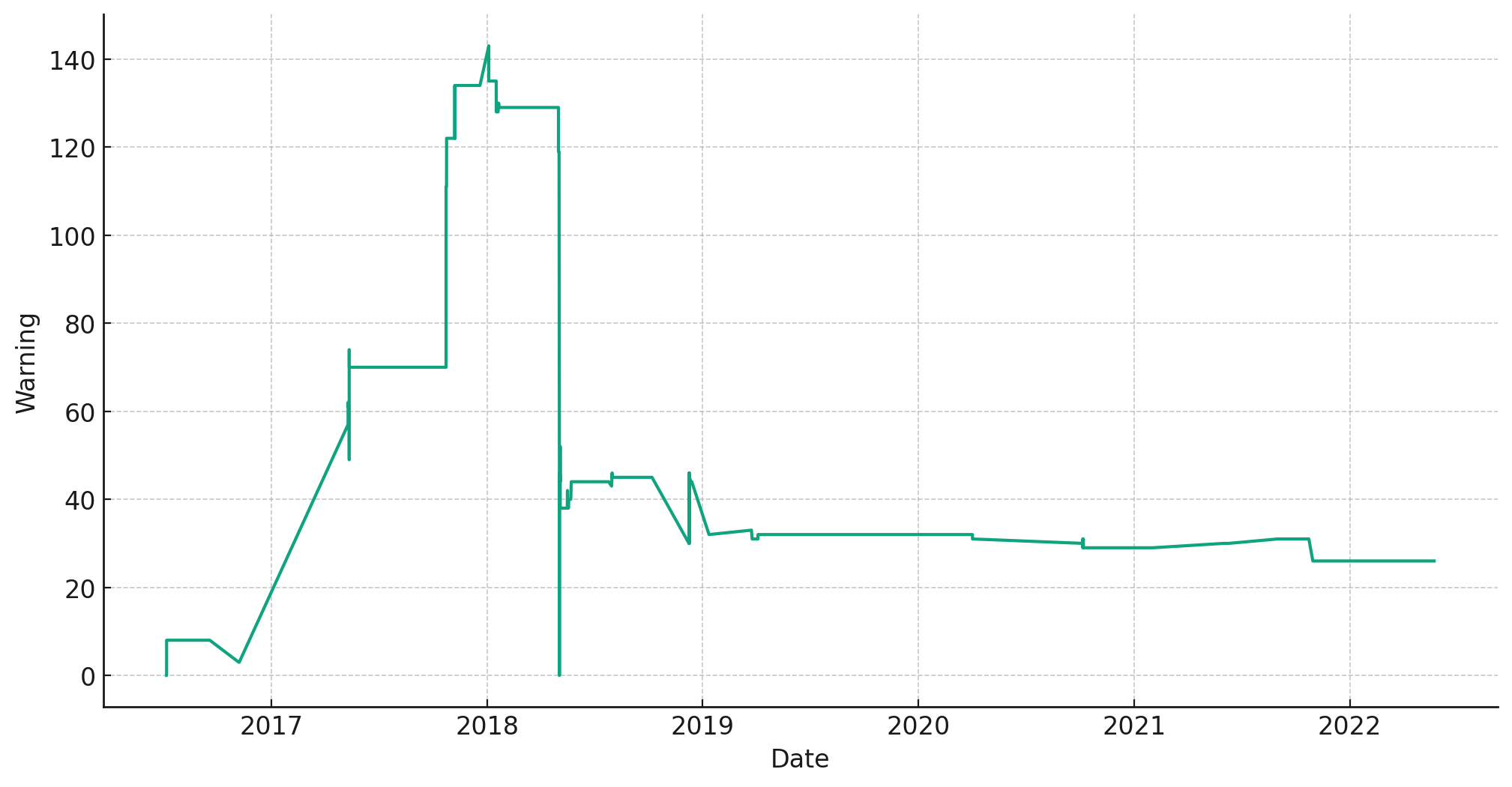}    & \includegraphics[width=0.45\textwidth]{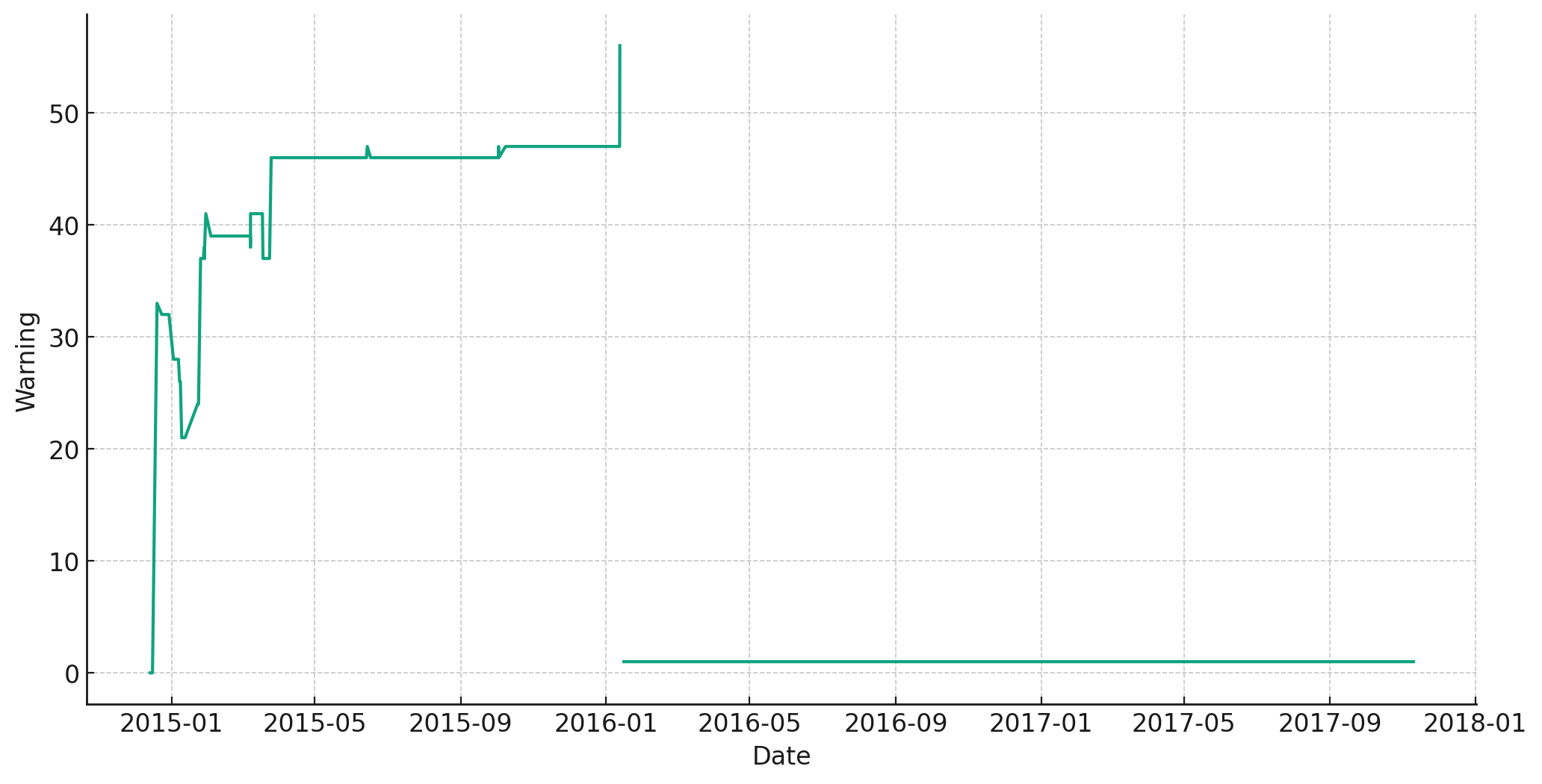}   \\
   Project RustFix & Project Log   \\
 \includegraphics[width=0.45\textwidth]{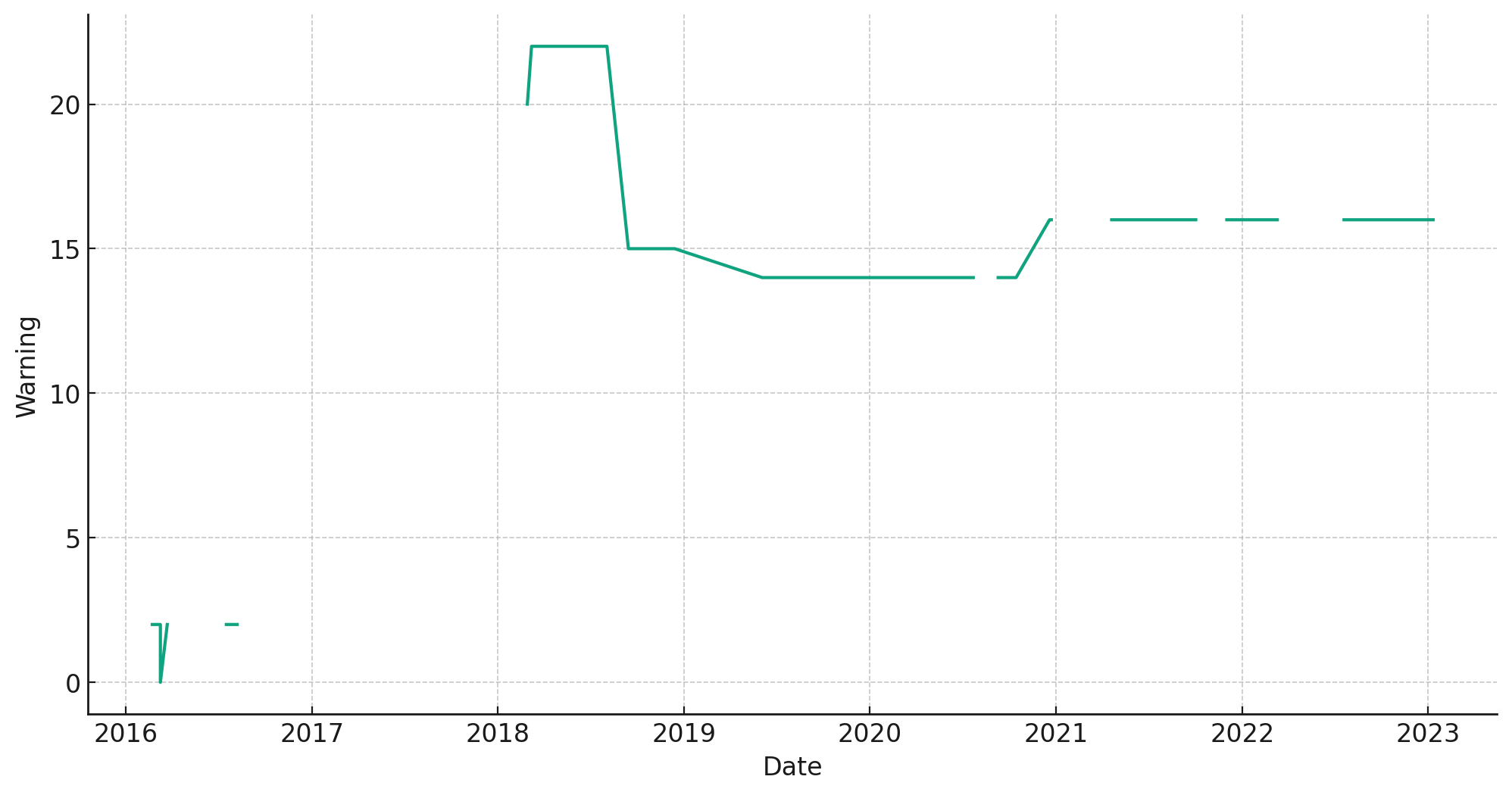}    & 
 \includegraphics[width=0.45\textwidth]{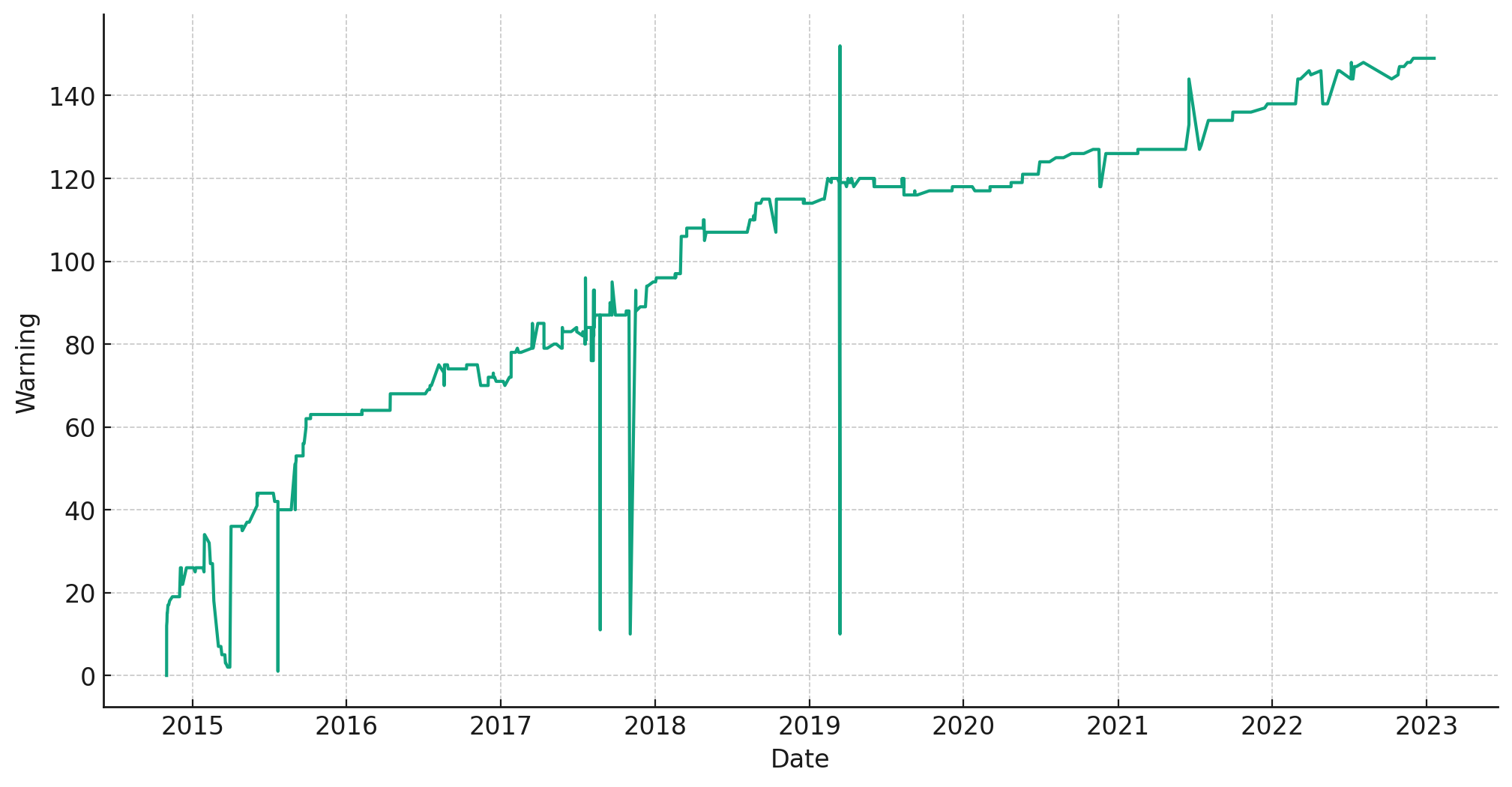}  \\
 Project Git2-Rs & Project CC-Rs
 \end{tabular}

% Warning trend over time
 \caption{Evolution of warnings over time for four typical Rust projects: RustFix, Log, %RustClippy, 
	Git2-Rs,
	and CC-Rs.
% 	are taken from the
% 	official rust-lang organisation. 
Horizontal axes show the actual date of commit activities; vertical axes show the number of warnings. While three projects (RustFix, Log, and Git2-Rs) have made commendable efforts in reducing Clippy warnings, CC-Rs has chosen to overlook most warnings.}
 \label{fig:plots}
 \end{figure*}

As observed in Figure~\ref{fig:plots}, the number of warnings in RustFix, Log, and Git2-Rs decreases significantly over time (The detailed data is in the supplementary.), indicating that these projects have likely adopted Clippy for gate-keeping practices and addressed many warnings, leading to a relatively stable warning count. However, the number of warnings in CC-Rs has exhibited an upward trend over the years, indicative of possible neglect in addressing Clippy warnings.

% In this case, however, the density per KLOC
% is not as low as the library Git2-Rs that depends on it. 

% \paragraph{Rust-Lang}
% Unlike the Crates/IO where majority of the projects are contributed by a wide
% community, the official Rust projects hosted by `rust-lang` may exhibit a
% different picture. There is not much point to compare absolute warning numbers,
% because of the varying size of the projects in terms of LOC; however, it is
% still interesting to see how dense these warnings are distributed.

% Across all evolution revisions of `rust-lang` projects we collected, we found
% in total 12,025,864 warnings. Figure~\ref{fig:rust-lang-warnings} shows the
% numbers of top-20 warning types amongst the \rustConventionRules{} Rust
% conventional lint rules. Amongst them, the {\tt unwrap\_used} rule also ranks
% no. 4 high frequency in the occurrences.

% \begin{figure}[th]
% \includegraphics[width=\columnwidth]{rust-lang.png}
% \caption{Top 20 types of warnings in the evolution of Rust projects (rust-lang), {\tt unwrap\_used} ranks No. 4  
% \label{fig:rust-lang-warnings}}
% \end{figure}

% \paragraph{Research prototypes}
% 这个图需要变与陈述相符合and清晰化；可以把下面的itemize重新总结成几句话
% Figure~\ref{fig:plots} shows a few representative Rust projects we studied in
% this work: 

{\bf Brief Answers of RQ1} {\it
The analysis of over 8 million warnings from the crates.io code landscape reveals a significant imbalance in the distribution of warning types. Additionally, the longitudinal study highlights that while some projects have made commendable efforts in reducing Clippy warnings, others have chosen to overlook these warnings, raising questions about the adoption of best practices.} %in rust development
% many warnings types also exist for more official `high-quality` Rust projects. 
%Among these, the {\tt unwrap\_used} warning stood out as one of the most frequently encountered, warranting a comprehensive investigation into its characteristics.
%\end{tcolorbox}

\section{Developers' Feedback to Clippy}
After discovering the persistence of projects like CC-Rs, which experiences a continuous rise in warnings without mitigation and imbalanced warning distributions on crates.io projects, we are eager to explore the factors that hinder the widespread adoption of Clippy. Therefore, in this section, we conducted a survey with aim to answer this research question:

\textbf{RQ2. User Survey} {\it Is Clippy considered a useful linter in real-world projects? What are the key factors that obstruct the widespread adoption of Clippy? }

% \begin{research}[User Survey] 
% \end{research}

\subsection{Experiment Design}
\paragraph{Participant selection} We distributed surveys to developers from four different categories, ranging from newcomers to Rust programming to experienced Rust community leaders. The survey was specifically sent to developers in two universities, individuals working in companies, and Rust community experts. A time period of 7 days was provided to collect their valuable feedback. Each respondent was requested to fill out a separate copy of the survey. They were allowed to use online search tools to find answers to the survey questions, with the exception of questions in RQs Set 2. It was also acceptable for participants to skip any questions they found difficult.

% We send surveys to four kinds of developers from newcomers for rust to rust community leaders. Specifically, the survey is sent to two universities, developers in the company, and rust community experts.  We leave 7 days to collect their feedback. Each person should fill in a separate copy; and can use online search tools to look for the answers to the survey questions excepted RQs Set 2. It’s ok to skip some questions for cannot understand questions. 

% 发了多少分，注意后面补充写The response rate is .

% Design the experiments:

We carefully designed five kinds of questions as follow~\footnote{The detailed questions and answers can be found in the supplementary materials.}.
\paragraph{RQ2.1}
First, we requested participants to assess their current proficiency in the Rust programming language on a scale ranging from 1 ("Beginner") to 10 ("Expert"). Subsequently, we devised a set of four questions to gather basic information about their familiarity with clippy. Specifically, we inquired whether they had prior experience using clippy and, if so, we further investigated the frequency with which they incorporated clippy into their development workflow, categorizing their responses into daily use, occasional use, or no usage. Moreover, participants were prompted to select other linters they have utilized, apart from clippy, from a list of four popular linters: rustfmt, cargo-geiger, MIRAI, and none of the mentioned linters. Additionally, we sought insights into the motivations behind using Clippy, offering respondents a choice of three possible reasons: adhering to institutional regulations, improving the idiomatic nature of their code, or enhancing the safety of their code.

% Subsequent questions (RQs Set 1-4) were presented, and participants were encouraged to respond to these inquiries if they had prior experience using Clippy.

\paragraph{RQ2.2}
To explore developers' ability to intuitively adhere to Clippy linting rules, we meticulously select 18 pairs of code snippets from the Clippy lints website~\cite{clippy-lints}. Within each pair, one code snippet triggers a warning from Clippy, while the other remains warning-free. Participants are prompted to identify or select the snippets that generate warnings. These examples were extracted from nine distinct warning types (such as Clippy::style and Clippy::correctness) available on the clippy website. To streamline feedback collection, the problem categorization at odd-numbered positions (e.g., 1 and 3) placed warning-inducing code under Option A, while for even-numbered positions (e.g., 2 and 4), code resulting in Clippy warnings was placed under Option B. To maintain impartiality, participants were explicitly advised against consulting search engines like Google. Moreover, we informed them that the code order had been thoughtfully shuffled to eliminate any predictable patterns, thereby avoiding situations where warning-inducing code consistently preceded warning-free code. 

\begin{figure}[h]
\centering
\begin{minipage}{.43\linewidth}
  \includegraphics[width=\linewidth,height=.8\linewidth]{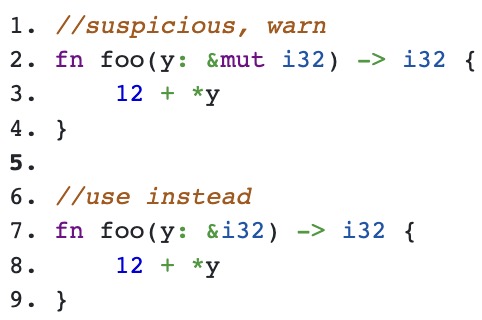}
\end{minipage}
% \hspace{.05\linewidth}
\begin{minipage}{.54\linewidth}
  \includegraphics[width=\linewidth,height=.8\linewidth]{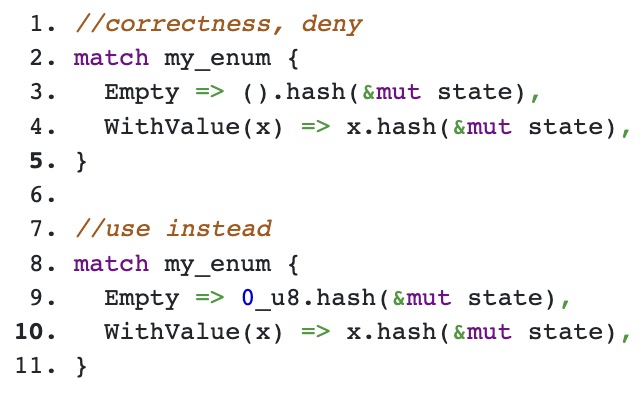}
\end{minipage}
\captionof{figure}{Code pairs with the above snippets triggering the warnings "needless\_pass\_by\_ref\_mut" (left figure) and "unit\_hash" (right figure) when analyzed by clippy. The below codes in both figures are the recommended replacement without warnings. }
\label{img:warn_examples}
\end{figure}

As depicted in Figure~\ref{img:warn_examples}, our survey comprises meticulously selected code pairs for rigorous evaluation. In the left figure, the above code snippets (Lines 2-4) trigger the "needless\_pass\_by\_ref\_mut" warning, categorized under Clippy::suspicious type. This specific rule contends that reducing the usage of \textsf{mut} may enhance opportunities for parallelization. Conversely, the right figure showcases the "unit\_hash" warning, categorized under Clippy::correctness type, signifying that hashing a unit value (e.g., ()) serves no purpose, as the implementation of Hash for () is a no-op. To effectively address these warnings and optimize overall code quality, clippy advises adopting the below codes, presented in both figures (Lines 7-9 in the left figure and Lines 8-11 in the right figure), as suitable replacements.

\paragraph{RQ2.3}
In this investigation, we explored the actual warnings encountered by developers from clippy during their daily projects. To gain insights into their clippy configurations, the benefits derived, and the impact on code quality, we presented a carefully crafted set of questions. We initiated the investigation by asking about the most frequently configured or enabled warning types in clippy, with the explicit aim of identifying specific instances. Additionally, participants were encouraged to candidly disclose their typical clippy configuration approach, encompassing four options such as utilizing existing presets, automatically generated configurations, minimal setups, or other customized options. Furthermore, developers' preferences regarding "warn" or "deny" settings when configuring Clippy were examined, unveiling their risk tolerance and treatment of warnings during development. By encouraging participants to share instances where clippy assisted in identifying potential issues or enhancing code quality, we sought to showcase the practical advantages of employing clippy. Moreover, participants were solicited to quantitatively rate clippy's overall effectiveness in improving code quality on a 5-point scale. Through exploration of the presence of similar warnings or errors reported by linters in other programming languages, we sought to ask them to identify common challenges across diverse language ecosystems and the transferability of experiences. Lastly, participants were advised to elucidate specific types of warnings or errors unique to clippy when juxtaposed with languages like C/C++, unveiling distinct language-specific characteristics in code analysis. These inquiries collectively provided valuable insights into clippy's role in enhancing code quality and its significance within the broader landscape of software development practices.

\paragraph{RQ2.4}
To comprehend how developers approach warnings in their coding practices, we prepared a set of concise questions to elucidate their perspectives and interactions with clippy's warnings. Firstly, we inquired about their typical response when encountering a warning from clippy. Participants were asked to select the most applicable option, which includes enabling the warning, manually addressing the warning without utilizing clippy's suggestions, using clippy check as a reference while fixing the issue manually, employing a different tool or linter to address the warning, applying clippy's autofix feature to automatically resolve the warning, or choosing to keep the warning as is. Additionally, we sought participants' attitudes towards fixing warnings prompted by clippy. Participants were invited to select all that applied from a list of sentiments, such as finding the warnings annoying, unhelpful, useless, unnecessary, distracting, or helpful. Moreover, we explored the helpfulness of clippy's autofix suggestions based on participants' experiences. Those who found the suggestions helpful were encouraged to provide specific instances. Furthermore, participants were asked to indicate the number of warnings that still required manual addressing even after using clippy's autofix feature. Additionally, we inquired about instances where clippy produced false positives or false negatives, aiming to identify any challenges developers may have faced in this regard. Participants were invited to provide examples to illustrate their encounters with such cases. Lastly, we sought participants' ideas on any aspects of clippy that they found confusing or difficult to understand, to gain insights into potential usability issues.

\paragraph{RQ2.5}
To collect feedback for clippy designs and contribute to the Rust community, we designed a survey encompassing a concise set of questions. Firstly, we sought participants' perspectives on the user-friendliness of clippy's warning messages and information, rated on a scale of 1 to 5. Next, we inquired about aspects of clippy that participants appreciated the most. Participants were asked to select all that applied from a list of options, including security, style guide (coding conventions), static analysis, performance improvements, and following optimal practices. Additionally, we investigated whether participants had contributed to clippy's warnings by submitting new lint suggestions or improvements, aiming to gauge the extent of community involvement. Moreover, we invited participants to share their insights on improvements or additional features they would like to see in clippy, providing valuable suggestions for further enhancements. Furthermore, participants were asked to rate on a scale of 1 to 5 how easy it was for them to set up and configure clippy in their development environment, offering insights into user experience and ease of adoption. We also sought feedback on the responsiveness and helpfulness of the clippy community in addressing issues or providing support, aiming to evaluate the level of community engagement. Participants were invited to share any challenges, limitations, or difficulties they encountered in using clippy, offering opportunities for improvement and addressing potential pain points. Lastly, we inquired whether participants would recommend clippy to other developers and the reasons behind their recommendation or reservations.

\subsection{Results Analysis}
In this section, we present the findings and discussions based on the feedback answers received. The detailed responses can be found in the supplementary materials accompanying this paper.

Totally, we gathered a total of 31 responses from developers with varying levels of expertise in rust programming, as self-assessed by the participants. 
% The response rate is ...
The feedback received from these 31 participants was deemed sufficient in achieving theoretical saturation, supported by findings from previous studies with participant numbers of 20~\cite{johnson2013don}, 18~\cite{layman2007toward}, and 15~\cite{tomasdottir2017and}. Notably, the participant group comprises eleven Master's students, two experts from the Rust community, and eighteen Rust developers employed in companies. In order to facilitate subsequent explanations and ensure the anonymity of the respondents, we will assign numbers to the respondents and use ``P'' as a prefix to indicate the order of each respondent. Next, we proceed to elucidate our analysis for five questions (RQs Set 0-4), based on the provided feedback responses.

% To facilitate our analysis, we made minor adjustments to some of the responses where necessary. Firstly, five participants's answers for RQs Set 1-4 are left blank because they had never used clippy before, so we excluded them from the feedback analysis pertaining to RQs Set 1-4.

\begin{figure}[h]
\centering
\includegraphics[width=\linewidth]{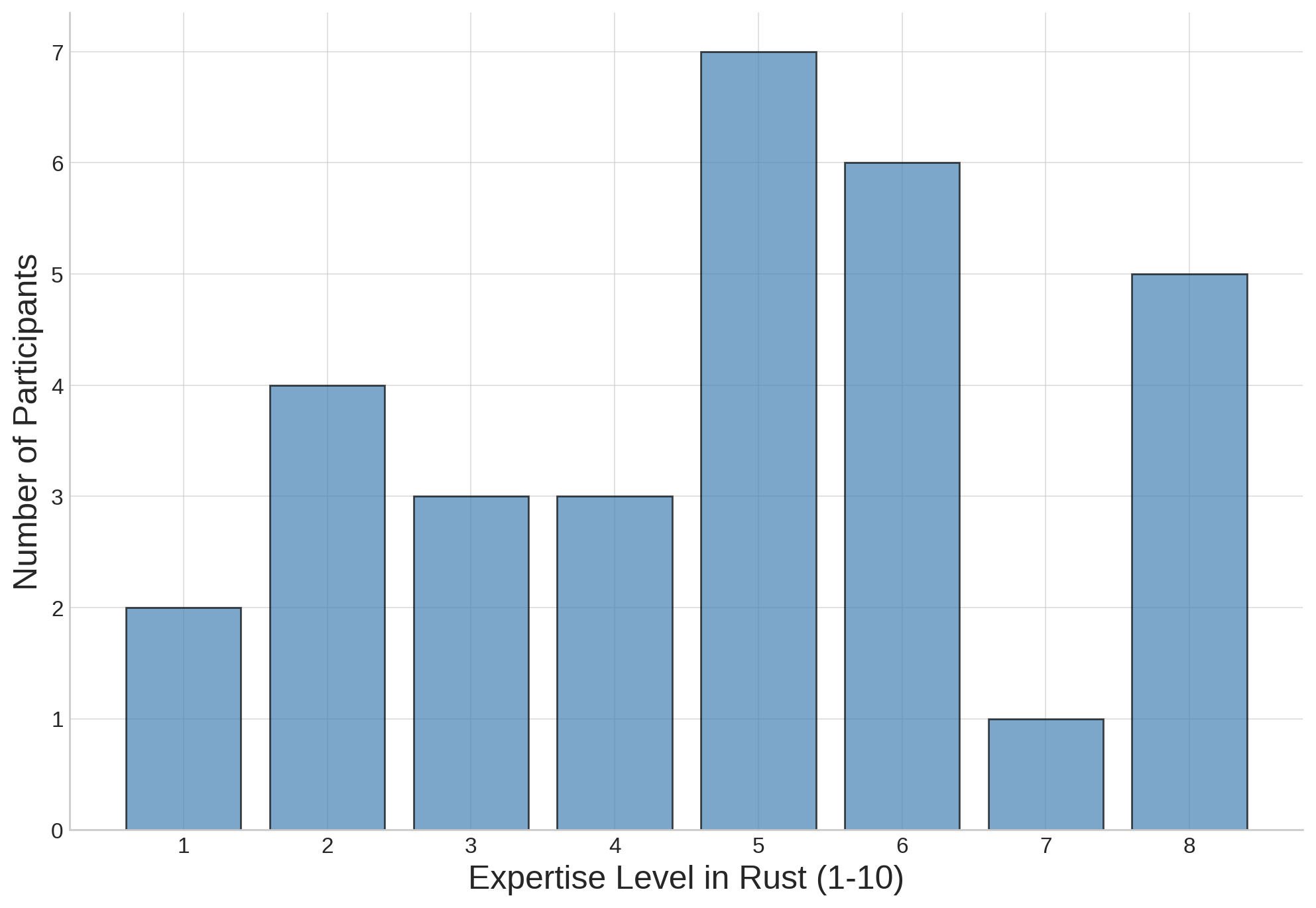}
\caption{The distribution of proficiency levels in Rust programming underscores the diversity of our participants' Rust programming skills.}
\label{img:rustLevelDistribution}
\end{figure}

\paragraph{RQ2.1}
Figure~\ref{img:rustLevelDistribution} presents the distribution of Rust developers across an array of proficiency levels. The proficiency in Rust programming is denoted on the x-axis, employing a scale that extends from 1 (representing a novice or beginner level) to 10 (indicating an expert level of proficiency). The y-axis portrays the participant count corresponding to each proficiency level. Most participants have an intermediate level of proficiency, predominantly between levels 4 and 6. This suggests that the majority of participants have some experience with Rust programming, but there is still a diverse range of expertise levels within the group.

The survey results indicates a high adoption rate among the participants with approximately 77\% (24 out of 31) utilizing Clippy. %This suggests that Clippy is widely accepted and used by Rust developers. 
The primary motivation for using Clippy, as stated by around 58\% of Clippy users, is to enhance their code style to align more with idiomatic Rust, while about 39\% employ Clippy to bolster the security of their code. These findings underline the significant emphasis placed on code quality by Rust developers. Examining the frequency of Clippy usage, we found a diverse range. About 29\% of Clippy users employ it occasionally, while close to 23\% use it daily. Around 13\% employ it weekly, while the remaining 35\% use it monthly or very rarely. In addition to Clippy, nearly 71\% of Clippy users also use \textsf{rustfmt} for code formatting, indicating a trend towards the use of multiple tools for code review within the Rust community.

% The average resonse time is ... after removed the useless feedback~\footnote{Five students did not used clippy before and did not provide feedback.}

\paragraph{RQ2.2}
% In this part, 
The purpose of the questions in this context is to gather insights into whether developers can naturally construct code that adheres to Clippy's lints.
Each participant was asked to answer 18 questions aimed at examining their understanding of different types of Clippy warnings. Participants identified by numbers [1-5, 11, 31] either declined to provide an answer or reported 'no warnings'. Their responses will be excluded from the subsequent statistical analysis. 

Based on our data analysis, the overall accuracy rate was 64\%, indicating that over half of the participants were able to correctly identify Clippy's warnings.
Interestingly, a higher proficiency level in Rust did not always correspond to a higher accuracy rate. Participants with a Rust proficiency level of 8 (participant numbers [10, 14, 15, 27]) had the highest accuracy rate (75\%, correctly answering 75\% of all questions), whereas those with a Rust proficiency level of 1 and 2 (participant numbers [20, 23, 24, 25]) had the lowest accuracy rate (44\%
%with approximately 44.44\% correct answers out of all questions
). This suggests that while Rust proficiency does play a role, it's not the only determinant of understanding Clippy warnings. Other factors, such as coding experience and habitual code review practices, might also be influential.

When we examined the response data based on the types of warnings, we found that participants had the highest understanding for Clippy::suspicious (Questions 2 and 17) and Clippy::cargo warnings (Questions 9 and 10), with average accuracy rates of 64\% and 61\%, respectively. In contrast, Clippy::complexity warnings (Questions 3 and 16) were the least understood, with an average accuracy rate of just 18\%.

% Despite the limited sample size, this study provides critical insights for improving the design and documentation of Clippy warnings, thereby enhancing code quality in the Rust community. Further research could provide a more comprehensive understanding of these findings and unearth additional factors influencing Clippy warning comprehension.

% One respondent explained why he chose the option. Foe example, for question 1, he attached "" %refer Usef's survey feedback

% 昨晚这些步骤去掉：
% 1. 要求ChatGPT编写代码用来发现数据中的模式
% 2. 当科学家将错误信息反馈给它，并要求其纠正错误时，它最终生成了可以用来探索数据集的代码。
% 3. 要求ChatGPT帮助他们制定一个研究目标。
% 4. 得出结论
% 然后，ChatGPT被要求在表格中总结关键发现，并编写整个结果部分。接着，他们逐步地要求ChatGPT撰写摘要、引言、方法和讨论部分。最后，他们要求ChatGPT完善文本。

% 需要对结果的可能的解释

\paragraph{RQ2.3}
% 考虑将某些结果表示成图
This part primarily focuses on how the participants configure and utilize Clippy, and their evaluation of Clippy's effectiveness in enhancing code quality. 

Participants had varying preferences for enabling warnings, with some like P7 opting for \textsf{clippy::style} and \textsf{clippy::suspicious} to avoid common mistakes, while others (like P30) chose \textsf{unwrap\_used} as ``This lint warns whenever .unwrap() or .expect() are used, encouraging safer error handling using .ok() or pattern matching.''. 
% A significant portion of participants (58\%) did not respond to this question. 

The majority of participants (48\%) used existing preset configurations for Clippy, with 26\% using an autogenerated configuration and 13\% opting for a minimal one. 
% We noticed that most developers (p7,..) in the company prefer to use the preset clippy configurations. students and experts prefer to use the automaticlly generated configurations.
Most participants (68\%) preferred 'warn' when configuring Clippy, while only 16\% chose 'deny'. Some participants provided examples of how Clippy helped catch potential issues, while others mentioned not encountering such cases due to limited experience in Rust programming. For instance, P30 shared: ``\textit{let opt = Some(5); let val = opt.unwrap(); // Clippy warning: use of `unwrap` on an `Option` value}'', and P31 stated: ``\textit{Not much as I haven't written much code in Rust myself.}'' The effectiveness of Clippy in improving code quality received a rating of 4 from the majority of participants (45\%) on a scale of 1 to 5. 

Regarding other programming languages, some participants noted similar warnings or errors in code checking tools. 
For instance, P19 stated: ``\textit{Yes. In c++ lint, there will be warnings such as writing explicit for constructors and noexcept for functions.}'' P7 revealed: ``\textit{Jetbrains Compiler. For example, both pycharm and IDEA will provide similar warnings, like the function is not used.}'' However, 48\% of participants stated they had not encountered similar types of warnings or errors for other languages. The views on the distinctiveness or depth of Clippy's warnings in comparison to other languages were split. Around 29\% of respondents thought they were not more distinct or comprehensive, while 39\% believed they were more unique or extensive, as evident by ``\textit{lifetime warnings (P26), clippy::missing\_safety\_doc provides guarantees for unsafe rust specifications (p27)}''.

% We noticed that one master student used \textbf{cargo 1.70.0 (ec8a8a0ca 2023-04-25)} to indicate warnings in cases where he was unsure which code option to choose. Notably, if the warning type is not explicitly specified in the file \textbf{clippy.toml} or command line~\footnote{Source: https://doc.rust-lang.org/clippy/configuration.html.}, Clippy may not consider the warning, especially if it does not belong to the default lint category. Throughout our survey, we assumed that Clippy had already enabled the warning types related to all options.

% Due to respondents' lack of precise knowledge regarding the specific warning types targeted in the questions, they may inadvertently respond with unrelated warnings. For example, for question 4, the P31

% This observation highlights a potential concern related to example code in clippy lints. To address this issue effectively, lints developers should ensure that the example code provided does not include warnings for other lints and remains specific to the particular warnings under consideration.

% Most participants rates for clippy helps them to improve the code quality. One participant complains that the clippy report ``too many not-so-important alarms'', which is also reflected by another survey (500 warnings for 50 lines)

\paragraph{RQ2.4}
The answers for this collection outlooks on how users interact with Clippy's warnings. Participants demonstrated diverse strategies for handling Clippy's warnings, with a notable portion favoring the "Check with Clippy" and "Fix with Clippy", indicating a reliance on Clippy's advice and autofix feature for resolution.

In terms of the helpfulness for fixing warnings message with Clippy, the majority of (74\%) respondents found it to be helpful. While a considerable 48\% acknowledged Clippy's auto-fix functionality valuable, a noteworthy 29\% perceived them as less effective. Especially, some said that ``\textit{Most of the time, autofix is not really useful as it concerns very few lints. (P11)}'' and ``\textit{it's not applicable to all warnings and many issues will still require manual intervention, particularly those that involve complex logic or potential design changes. (P30)}''. Hence, the development of automated techniques to assist in resolving Clippy warnings becomes crucial.

Surprisingly, a lot of (42\%) participants complained that Clippy produce relatively many and frequent false positives. for instance, ``\textit{For instance, it may warn about unused variables in macros that do get used when expanded, or unnecessary parentheses in certain complex expressions.} (P30)''. Additionally, some respondents mentioned the confusing or challenging parts of Clippy, like unclear lint explanations and examples (P27, P12 and P30), and so many warnings on only a few codes (P6). 

% Regarding the aspects of Clippy that are confusing or hard to understand, 23\% indicated they have, such as "many warnings lack suggestions, can be inaccurate and distracting." 

% Annoying clippy warnings, to escape this, developers can enable/ allow specific types of warnings, or they don't configure this warning if the warning type does not belong to the default lints category. 

% ---------------

% Three participants (No .., .) complained about the False positives

% ··many warnings lack suggestions, can be inaccurate and distracting‘’

% ``Quite some clippy lint warnings are specific to some unique features in the Rust language, which need better explanation for newbies.''

% One response is that ``\textit{Clippy's autofix feature can be very helpful. However, it's not applicable to all warnings and many issues will still require manual intervention, particularly those that involve complex logic or potential design changes.}'' (P30). Our research addresses the concern by proposing three feasible solutions to rectify warnings beyond Clippy's automated fixes, which will be elaborated in Section~\ref{sec:case_study}. The term 'Clippy autofix' denotes the application of \texttt{cargo clippy --fix}.
% ---------------

\paragraph{RQ2.5}
The survey gathered valuable insights and feedback on Clippy designs. The majority of respondents found Clippy's warning messages and prompts to be user-friendly (58\% rated 4 or 5). Users appreciated Clippy for adhering to optimal practices (53\%) and style guides (35\%). Only four participants contributed to Clippy's warnings. Desired improvements to Clippy include providing better solutions for fixing warnings (P8), supporting dynamic check (P16), and reducing false positives (P30). Setting up Clippy was generally considered easy (58\% rates 4 or 5). Some participants (like P11 nad P12) praised the Clippy community as responsive and helpful, while others (like P8) had not engaged with it. Almost all participants (except empty replies and one complain about the complex configuration (P18)) reported no challenges with Clippy. All replies indicates that they would recommend it to others except for one thought it is not useful as rustc's built-in lints (P6). 

\subsection{Discussion and Implication}
% In summary, our findings reveal a significant variation in the understanding of Clippy warnings among Rust developers. The variation seems to be influenced by a combination of factors, including Rust proficiency level, coding experience, and familiarity with specific warning types.
Our survey findings reveal new implications to three beneficiaries:
% for Rust developers, researchers, and Clippy community, which we clarify as follows.

\paragraph{Rust developers}
A considerable portion of surveyed developers highly value Clippy for its adherence to optimal practices and style guides. Thus, we strongly suggest Rust developers consider incorporating Clippy into their projects. The auto-fix feature of Clippy remains beneficial, and it is advisable to carefully review warning fix suggestions to address potential false positives and improve code quality. Clippy's warnings provide distinctive insights not commonly found in other programming languages. Furthermore, alongside Clippy, a majority of developers also utilize \textsf{rustfmt} for code formatting needs.

\paragraph{Researchers}
Respondents strongly voiced their desire for improved solutions to address warnings, underscoring that auto-fix functionality can only handle simple errors. Researchers should explore novel techniques to assist users in effectively resolving warnings, with particular attention to addressing the most frequently occurring warnings as highlighted in Section 3. Additionally, the prevalence of false positives in Clippy's warnings indicates the necessity of developing supplementary tools to verify the accuracy of linting outcomes. Researchers have the opportunity to concentrate on creating complementary tools that enhance Clippy's analysis, thereby elevating its reliability and precision.

\paragraph{Clippy community}
The feedback provided underscores the paramount significance of comprehensive and lucid documentation in augmenting user comprehension. As creators of lints, it becomes imperative to ensure that the documentation is exhaustive, logically organized, and accompanied by illustrative examples. Moreover, being responsive to user inquiries and incorporating their useful suggestions can lead to continuous improvements, enhancing the overall effectiveness of Clippy. Providing improved solutions for fixing warnings can further boost user satisfaction and usability. Additionally, the prevalence of false positives in Clippy's warnings highlights the necessity for supplementary tools aimed at corroborating the accuracy of linting outcomes. Lastly, it is crucial that each incorporation of a lint undergoes meticulous examination and adheres to rigorous review standards.

{\bf Brief Answers of RQ2} {\it Clippy is regarded valuable for real-world projects, especially for adhering to coding practices and style guides. However, its widespread adoption is mainly hampered by false positives and less effective warning auto-fix techniques.}

\section{Case Study to Fix Clippy Warnings}
\label{sec:case_study}

\textbf{RQ3. Auto Fixes} {\it How to fix high-frequency Clippy lints to reduce the warning density for benchmarks?}

The prevalence of innumerable unconcerned warnings generated by the Clippy presents a vexing challenge for software developers. This issue is further exacerbated for developers employed in large corporate organizations, where compliance with institution-regulated lints is an obligatory requirement. Such adherence holds profound importance as it ensures the viability of the produced code for integration into commercial products and facilitates ease in subsequent code maintenance and uniformity in coding styles. Given these circumstances, it becomes of utmost significance to assist developers in promptly resolving these warnings. By doing so, the development process can be accelerated, enabling quicker iterations and enhanced productivity in software development endeavors.

% Especially, some said that ``\textit{Most of the time, autofix is not really useful as it concerns very few lints. (P11)}'' and ``\textit{it's not applicable to all warnings and many issues will still require manual intervention, particularly those that involve complex logic or potential design changes. (P30)}''. Therefore, it is essential to develop automatic techniques to help fix Clippy warnings.

% Many respondents mentioned that ``Provide better solutions for fixing warnings'' (p8), ``auto-fix is generally used for simple errors. Most of the problems encountered now do not have automatic repair functionality.''(P15), so that we propoed (也可以put to the motivation of sec 5)

% How can we improve the performance of Clippy?

Based on the findings from our survey, Clippy's auto-fix feature proves helpful in resolving code issues. However, it is essential for developers to be cognizant of its limitations. The Clippy solely asserts that suggestions related to lints tagged as \textsf{MachineApplicable}~\cite{applicability} should be automatically implemented, while the majority of warnings pertain to other categories, namely \textsf{MaybeIncorrect}, \textsf{Unspecified}, or \textsf{Unresolved}. Consequently, developers are required to manually verify the correctness of the provided solutions in these cases. As a result, Clippy's solutions should be viewed as supplementary rather than comprehensive. Furthermore, our observations indicate that despite Clippy automatically removing a warning, it may lead to the emergence of another new warning. For instance, as illustrated in Figure~\ref{fig:example}, the elimination of an \textsf{unwrap\_used} warning results in the appearance of another warning, \textsf{expect\_used}.

\begin{figure}[h]
\centering
\includegraphics[width=\linewidth]{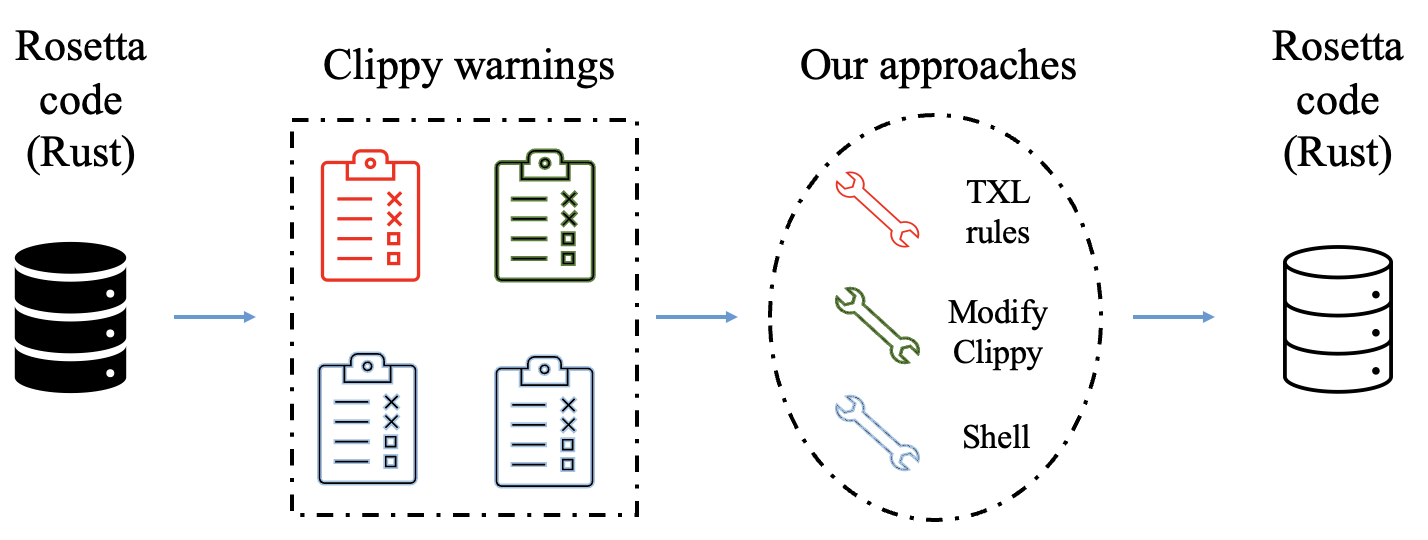}
\caption{Three methods to effectively mitigate 
%the occurrence of 
the four most frequent types of warnings in the Rosseta Rust dataset.}
\label{img:warningFix}
\end{figure}

In this section, as shown in Figure~\ref{img:warningFix},
we present three viable approaches to effectively mitigate the occurrence of the four most frequent types of warnings found in the Rosseta Rust dataset, as identified in CRustS~\cite{ling2022crusts}. The first approach involves the utilization of rule transformation, wherein we manually constructed TXL rules to reduce the \textsf{arithmetic\_side\_effects}~\cite{arithmeticSideEffects} warnings. Subsequently, we propose a modification to Clippy's existing functionalities, aiming to seamlessly integrate a warning fix operation directly into Clippy itself to fully eliminate \textsf{default\_numeric\_fallback}~\cite{defaultNumericFallback} warnings. Thirdly, we devised shell commands to alleviate the two common types of warnings: \textsf{undocumented\_unsafe\_blocks}~\cite{undocumentedUnsafeBlcoks} and \textsf{missing\_debug\_implementations}~\cite{missingDebugImplementation}~\footnote{ This lint belongs to rustc compiler's lints.}

\subsection{Experiment Design}
%  论文陈述事实 结论和claim对的上 可以不用夸大论述  所有数据的真实性都需要让Luca从新跑 放到supplementary里面 -> 

% 我们的目标就是先消除最高频的warning，清零不支持
      
%  同样的方法用到crateIO上，warning能消除的数字 _ KLOC -> _ KLOC

The dataset under consideration is derived from the Rust variant of the Rosetta dataset, transformed via the CRustS methodology as presented by Ling et al. (2022)~\cite{ling2022crusts}. There exist several Rust coding convention and style guidelines. From the
official Rust API
Guidelines\cite{api-guideline}, the official
Rust Style Guide\cite{fmt-rfcs},
the Rust Unsafe Code
Guidelines\cite{unsafe-code-guidelines},
French ANSSI Safe Rust Code Guidelines
\cite{coding-guidelines}, and the
guidelines adopted by influential large-scale Rust projects such as Facebook
Diem, Apache Teaclave, PingCAP and Google Fuchsia, we extracted 33 clippy rules
as mandatory. 
% \todo{Prof. Yu check} OK

% draw a graph to show warnings frequency on rosseta dataset
\begin{figure}[h]
\centering
\includegraphics[width=\linewidth]{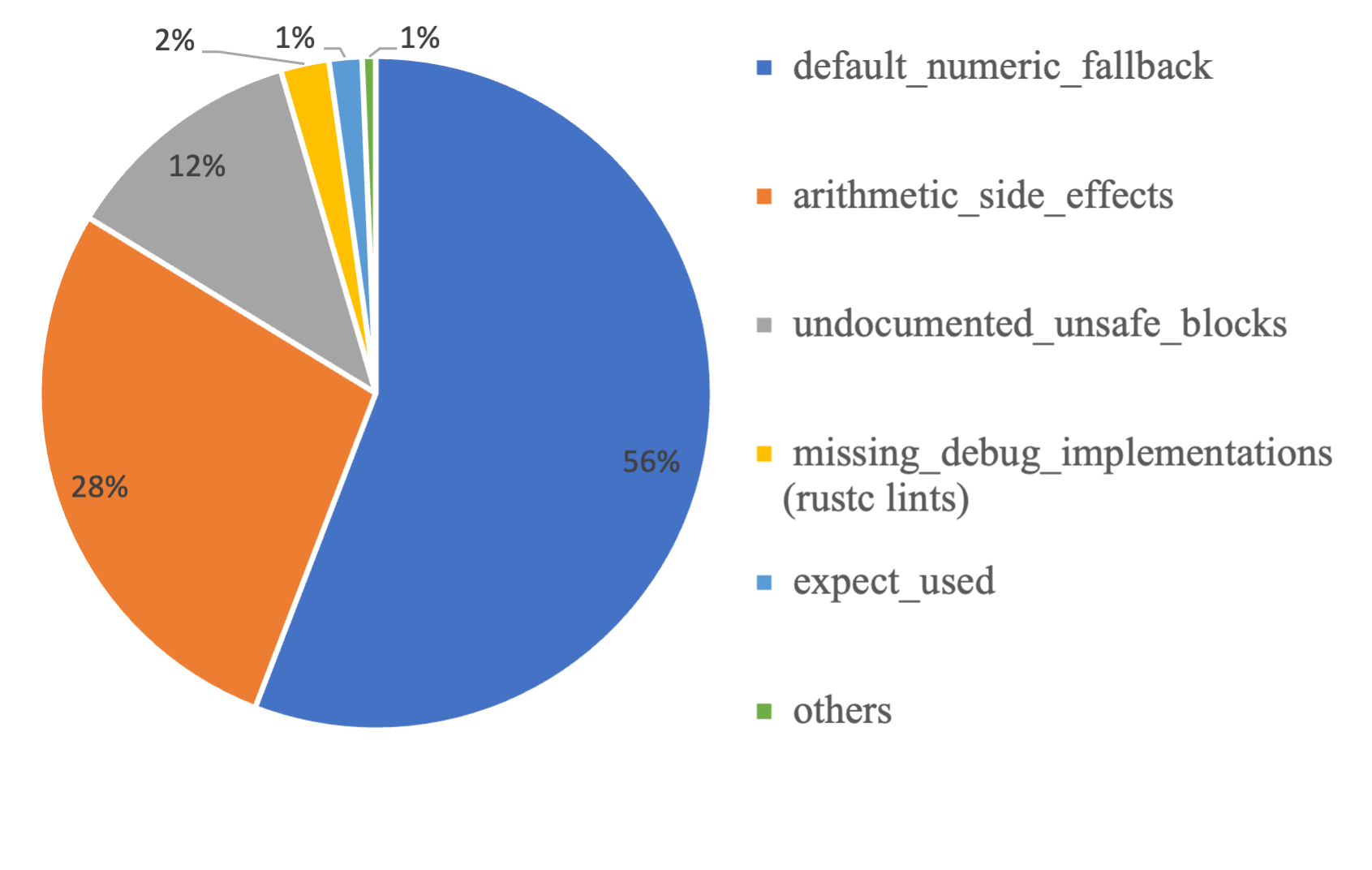}
\caption{Warnings distribution on CRustS Rosseta dataset}
\label{img:lint_distr}
\end{figure}

Upon applying Clippy's rules to our dataset with the auto-fix option enabled, we present the distribution of the resulting warnings in Figure~\ref{img:lint_distr}. The data reveals that the four most frequent warning categories are `default\_numeric\_fallback`, `arithmetic\_side\_effects`, `undocumented\_unsafe\_blocks`, and `missing\_debug\_implementations`. Of particular note is the `default\_numeric\_fallback` category, which represents a significant 56\% of the total warnings. Consequently, our objective is to eliminate these four prevalent warning types.

It is noteworthy that the auto-fix feature provided by Clippy, invoked via the `cargo clippy --fix` command, is unable to rectify these warnings within our dataset. To overcome this limitation, we have developed three unique strategies for their resolution. We have observed in some projects that developers employ the `allow` attribute to evade these warnings. However, we do not consider this approach as a valid resolution to the warning issue. Instead, this method merely suppresses the display of warnings without addressing the underlying problems.

For the purposes of our warning fix task, we have defined correctness as the absence of the same warning when executing ``\textsf{cargo clippy --warning name}'' following the implementation of our proposed solutions. Next, we illustrate our three different approaches with more details.

% The overall workflow is drawn in Figure~\ref{img:warningFix}. 

% To make the generated rust programs safer, previous work conducted studies to reduce the unsafe ratio in transformed rust programs~\cite{}.

% why clippy cannot provide the auto-fix, challenges

\subsection{Three Approaches}

\subsubsection{Clippy Modification}
\textsf{default\_numeric\_fallback} is the top1 warning in the Rosetta dataset. Elimination of this warning will significantly reduce the percentage of warnings to the code lines. For example, the pattern of these lints is as simple as Figure~\ref{img:9}. Meanwhile, the hint for this warning is precise enough to fix the warning without worrying side effect caused by the modification. Based on these facts, we modified the source code of Clippy by replacing the tag "MaybeIncorrect" of this lint with "MachineApplicable" to enforce it to strictly follow the hint to automatically fix the warnings. The final result proves this kind of fix is implementable and without side effect for further warning elimination. 

\begin{figure}[h]
\begin{lstlisting}[style=icse-style,escapechar=\%]
error: default numeric fallback might occur
 --> src/main.rs:112:16
  |  return 0;
  |         ^ help: consider adding suffix:`0_i32`

\end{lstlisting}
\caption{Warning occurrence of default\_numeric\_fallback\label{img:9}}
\end{figure}

% \begin{figure}[h]
% \centering
% \includegraphics[width=\linewidth]{dnfwarn.png}
% \caption{Warning occurrences of default\_numeric\_fallback}
% \label{img:9}
% \end{figure}

\subsubsection{Rule Transformation}
% "arithmetic\_side\_effects" is the top1 warning in the Rosetta dataset. Elimination of this warning will significantly reduce the percentage of warnings to the code lines. \todo{Haitao recheck}
The warning patterns under the \textsf{arithmetic\_side\_effects} category pose a considerable challenge for automatic fixing by Clippy. The reason is that these patterns exhibit a wide range of situations and contexts, as evident from the examples shown in Figure~\ref{img:7}. As a result, Clippy's automated approach finds it difficult to handle the diverse nature of these warnings effectively.

\begin{figure}[h]
\begin{lstlisting}[style=icse-style,escapechar=\%]
error: arithmetic operation that can potentially result in unexpected side-effects
 --> src/main.rs:58:13
  58  |  i += 1; 
      |  ^^^^^^

error: arithmetic operation that can potentially result in unexpected side-effects
--> src/main.rs:30:15
  30  |  c[(1+j) as usize] = 1; 
      |    ^^^^^

error: arithmetic operation that can potentially result in unexpected side-effects
--> src/main.rs:19:13
  19  |  n /= p;
      |  ^^^^^^
\end{lstlisting}
\caption{Warning occurrences of arithmetic\_side\_effects\label{img:7}}
\end{figure}

% \begin{figure}[h]
% \centering
% \includegraphics[width=\linewidth]{asewarn.png}
% \caption{Warning occurrences of arithmetic\_side\_effects}
% \label{img:7}
% \end{figure}

In this intricate scenario, we have employed TXL to meticulously construct specific rules targeting subsets of this warning. As a result, we have curated a collection of TXL rules that work in unison to effectively eliminate this particular type of warning. Notably, our approach ensures the complete removal of these warnings without introducing new ones. It is worth mentioning that the hints provided by Clippy do not propose solutions akin to ours.

As illustrated in Figure~\ref{img:8}, these rules correspond precisely to the cases depicted in Figure~\ref{img:7}. For instance, the rule \textsf{eqAddFix} modifies \textsf{i += 1} to\textsf{i = i.wrapping\_add(1)}, showcasing the corrective nature of our rules. The comprehensive set of manually crafted TXL rules, designed to handle all analogous warnings stemming from the \textbf{arithmetic\_side\_effects}, is available in the supplementary materials. Moving forward, our objective is to explore automated methods for resolving these warnings. Three simple rules aim to solve three pattern of warnings, and importantly, they will not introduce error into the code or bring side effect to impact the result either.

\begin{figure}[h]
\begin{lstlisting}[language=Txl,style=boxed]
rule eqAddFix
    replace [Statement]
        varId [id] '+= N [integer_number];
    by
        varId '= varId.wrapping_add(N);
end rule
rule AddExpFix
    replace [Expression]
        varId [id] '+ N [integer_number]
    by
        varId.wrapping_add(N)
end rule
rule eqDivFix
    replace [Statement]
        varId [id] '/= N [integer_number];
    by
        varId '= varId.wrapping_div(N);
end rule
\end{lstlisting}
\caption{Some TXL rules fixing the arithmetic\_side\_effects % which fixes our example problem
\label{img:8}
}
\end{figure}

\subsubsection{Construct Script}
Regular expression script is the third way in this paper to fix the typical warnings. We provided two scripts for two warnings. For "missing\_debug\_implementations", the script will find all the keywords "struct" and "enum", then place the debug trait derive on top of it. For the "unsafe\_block\_missing\_comment", script will search the "unsafe" keyword, create a line above the whole unsafe block with a comment explaining that these unsafe codes were machine generated by c2rust. Notably, these two scripts will NOT introduce side effect when eliminates these two warnings.

\subsection{Result Analysis}

We collected the number of warnings~\footnote{The data related to our study are included in the supplementary materials.} using Clippy, the total lines of code, and computed the KLOC for three Rust variations of the Rosseta dataset: c2rust, CRustS, and the refactored data using our three automated approaches. The performance change statistics are presented in Table~\ref{tab:1}, which demonstrate the effectiveness of our warning fixes, reducing the number of warnings from 9704 in c2rust to 812 using our methods. Particularly noteworthy is the significant decrease in warning density from 195/KLOC in c2rust to an impressive 18/KLOC after applying our methods, already lower than the average density of 21 KLOC observed in crates-io Rust projects. We carefully examined 100 programs (around 1/3 of all programs) to ensure their functional correctness. As we move forward, we plan to expand this verification process to include more programs for a thorough examination.

% To ensure accuracy, we manually verified the correctness of 50 programs, and leave rigorous verification for more programs as future work.

Table~\ref{tab:2} presents how our approach had a significant impact on the top four common types of warnings before and after using the CRustS dataset. It was exciting to witness the substantial effectiveness of our approach in reducing warnings on the CRustS baseline, even though the dataset already had fewer warnings compared to c2rust. All four types of warnings experienced significant decreases. We observed that CRustS improved the safety of c2rust by sinking the "unsafe" on function identifiers into the function body, which reduced the ratio of unsafe functions but increased the ratio of unsafe blocks. As a result, the "undocumented\_unsafe\_blocks" warnings from CRustS (806) were more than those from c2rust (325). However, our script managed to greatly decrease this type of warning (5), further enhancing the safety of CRustS by alleviating warnings. Moreover, we successfully eliminated all instances of the \textsf{default\_numeric\_fallback} warnings, which showcases the inspiring effectiveness of our study.

% add percentage description

% Given the table
% Luca check the data
\begin{table}[t]
\begin{tabular}{|c|r|r|r|}
\hline
               & Warnings     & Lines & KLOC        \\ \hline
c2rust         & 9704         &  49791     & 195         \\ \hline
CRustS         & 6885         & 43863 & 156         \\ \hline
Our approaches & \textbf{812} & 43669 & \textbf{18} \\ \hline
\end{tabular}
\caption{Clippy-reported warnings data for three types of Rust codesets: c2rust, CRustS, and our approaches. Our methods significantly reduced warnings to only 18 KLOC, compared with 195 (c2rust) and 156 (CRustS).}
\label{tab:1}
\end{table}

\begin{table}[]
\begin{tabular}{|c|r|r|r|}
\hline
                                                                                         & c2rust & CRustS & \begin{tabular}[c]{@{}c@{}}Our \\ methods\end{tabular} \\ \hline
arithmetic\_side\_effects                                                                     &  2480      & 1919   & 682                                                       \\ \hline
default\_numeric\_fallback                                                               &   5789     & 3842   & 0                                                         \\ \hline
undocumented\_unsafe\_blocks                                                             &   325     & 806    & 5                                                         \\ \hline
\begin{tabular}[c]{@{}c@{}}missing\_debug\_implementations \\ *(rustc lints)\end{tabular} &  197      & 160    & 8                                                         \\ \hline
\end{tabular}
\caption{Numbers regarding the top four most frequently occurring warnings in three Rust codesets (c2rust, CRustS, and our approach). Our methods effectively reduced warnings, especially eliminating default\_numeric\_fallback. The initial trio of lints originates from the Clippy, whereas the final one stems from the rustc compiler.}
\label{tab:2}
\end{table}

% mention: warnings 总数> 四类
{\bf Brief Answers of RQ3} {\it %To summarize, 
We proposes three promising approaches to fix Clippy warnings efficiently. Notably, these strategies address the limitations of Clippy's auto-fix functionality, where certain warnings were previously left unattended or even introduced new warnings inadvertently. The three approaches involve rule transformation, direct integration of fix operations into Clippy, and the development of specialized shell script. Together, they offer effective means for developers to remove warnings and enhance the overall quality and reliability of Rust projects. It is worth noting that, to the best of our knowledge, we are the {\bf first} to tackle this specific issue. While our initial experimentation has successfully eliminated warnings in our dataset, future work will focus on validating the soundness and generalizability of these approaches across diverse Rust projects.}

% \section{Discussion}
\section{Threat to Validity}

% The survey design and time constraints may have limited the scope of questions, potentially impacting the comprehensiveness of the results. Future surveys could benefit from offering more individualized questions to capture a wider range of responses.

% Although we tell the participants not to search answers for RQs Set 2 (ie., to choose the code with warnings from 18 pairs of code snippets), they might do that. 

The questions in RQs Set 1 on the survey might not fully capture the intended scope of our study. It could be more beneficial to provide additional questions, allowing respondents to answer individually. However, due to the lengthy time required for completion, some respondents showed reluctance to participate. In the future, we plan to invite more Rust developers to take part in the experiments to enhance data representation. Our current results have largely indicated that respondents might not be entirely proficient in intuitively constructing Rust programs as regulated by Clippy lints. Moreover, the binary options of "A" or "B" in some questions have caused confusion for participants, leaving them unsure of how to respond. To address this, we will introduce a third option, "don't know," to offer respondents more clarity and flexibility in their answers.

% The questions for RQs Set 1 on the survey might not reflect what we want to test. INstead, it might be better that provide more questions to let respondents to individually answer. But that is due to the time cost is so long that a few repsondents are not willing to fill in our survey. IN the future, we consider invite more rust developers to complete this experiments. However, our results has almost reflect that the respondents are not adept in intutitively construct rust programs as clippy lints regulated. Also, the options only A or B so that some don"t know how to choose. IN the future, we will add the third option "don't know".

% We applied 50 lint rules to check warning distribution on Section 3, whereas we employed 30 rules to run on the Rosseta dataset.

We chose not to apply three warning fix approaches to the crates.io dataset for two main reasons. Firstly, the safety of the crates.io dataset has not been verified, and we preferred to work with a higher ratio of safety code, such as the Rosseta code generated from CRustS. Secondly, we observed that the two warning types we tackled, namely \textsf{default\_numeric\_fallback} and \textsf{arithmetic\_side\_effects}, held the first and third positions in the crates.io dataset. We plan to conduct future tests to evaluate the effectiveness of our three methods on the crates.io dataset.

% We applied three warning fix approaches on Rosseta code, which is generated from CRustS. We did not apply them on crates.io dataset. becasue two reasons: 1. the safety on crates.io dataset has not been verified. Instead, we prefer the higher ratio of safety code, like Rosseta code generated from CRustS. 2. we found that the four most common warning types we fixed also have high occurences percentage in the crate.io dataset. In the future, we will test the effectiveness of our three methods on crates.io dataset.

% c2rust只是c99标准 rosseta是c11标准的 我们只处理正确转换的 c99的结果

\section{Related Work}
\subsection{Empirical Studies on Rust}
Previous work has done some empirical studies regarding security issues in real-world Rust programs. Xu et al.~\cite{xu2021memory} surveyed 186 memory-safety bug reports to generate several major categories. Evans et al.~\cite{evans2020rust} studied the usage of unsafe code and indicated that the {\tt unsafe} keyword was used in less than 30\% Rust libraries, but the propagation of unsafe in the library’s call chains induced difficulty for the static check by compiler.  Qin et al.~\cite{qin2020understanding} empirically studied concurrency bugs in real-world Rust programs and revealed their impacts. Compared with their work, we also gain a better understanding of real-world Rust projects, but we focus on how clippy warnings are distributed among lint types, whether and how they can be fixed automatically.

% Errors and warnings are diagnostic feedback provided to programmers byautomated tools such as compilers and linters. Recognising their importance for software quality, responsible software engineering vendors mandate the use of lint rules for gate-keeping code commits. 

\subsection{Repairing Lint Violation}
% A linter flags the suspicious code snippets based on certain rules by emitting warnings or errors, either about the code convention checking or buggy constructs. 

Some studies proposed different solutions to remove certain kinds of linter violation cases. Phoenix~\cite{bavishi2019phoenix}, Getafix~\cite{bader2019getafix}, and SpongeBugs~\cite{marcilio2020spongebugs} repaired bugs and code smells reported by other static analysers. STYLER~\cite{loriot2022styler} and ESLint focused on formatting repair. Also, some tools~\cite{carvalho2020c, marcilio2020spongebugs} generate fix advice to facilitate developers to accept such suggestions to maintain code-bases. For example,  SpongeBugs~\cite{marcilio2020spongebugs} repaired 11 distinct rules violations detected by two code analysis tools (SonarQube and SpotBugs) for Java programs and automatically generated patching recommendations based on certain templates. 
% Moreover, specific compiler syntactical errors were repaired by using machine learning. Gupta et al.~\cite{gupta2017deepfix} designed Deepfix to build a language model to fix 27\% syntax errors in compilation for C programs, which was outperformed by TRACER~\cite{ahmed2018compilation} with 44\% repair percentage.
Whereas, Clippy rules considered in our work include more types of linting rules than the existing mentioned work, which only focuses on specific code smells or formatting rules. In other words, our warning patterns are more diverse and complicated than others. From what we know, we did the \emph{first} survey to show clippy warning distribution on all the Crates-IO Rust projects, and measured the effectiveness of three approaches for automated fixes on Rosseta dataset from CRustS. %As the Rust language is becoming popular and accepted by many projects, we call for more research attention on clipply warning repair.

% \subsection{Linter Investigation in Idiomatic Programs}

\subsection{User Study for Linters}
Numerous user studies~\cite{johnson2013don,tomasdottir2017and,habchi2018adopting,rafnsson2020fixing,tomasdottir2018adoption,christakis2016developers} have been conducted to explore users' experiences with linters. Johnson et al.\cite{johnson2013don} analyzed the reasons that impede software developers from using static analysis tools to find bugs. Tomasdottir et al.\cite{tomasdottir2017and} investigated the reasons and means by which JavaScript developers use linters. Habchi et al.~\cite{habchi2018adopting} studied the experience of adopting linters for addressing performance issues in Android apps. Our study, in contrast, aimed to gain feedback specifically related to Clippy from Rust developers with diverse levels of programming expertise.

% \subsection{Static Analysis for Rust}

\section{Conclusion}
In conclusion, our study has unleashed the significant impact of Clippy linter violations in real-world Rust projects. Through our investigation, we successfully identified the most frequently occurring warnings in all idiomatic crates.io projects. Additionally, valuable insights were garnered from the user survey, offering benefits to Rust developers, researchers, and the Clippy community alike.

Furthermore, our research has proposed three viable warning fix solutions that effectively reduced the occurrences of the top-four common warnings in the Rust version of Rosseta code stemming from CRustS. Moving forward, we aim to explore novel techniques to enhance Clippy's warning fix rate on a broader range of idiomatic projects. Additionally, we plan to enrich our user survey to gather further feedback specific to Clippy, fostering continuous improvement and enhancement of the linter tool.

\bibliographystyle{ACM-Reference-Format}
\bibliography{sample-base}

% \appendix

\end{document}